\documentclass[aip,jap,twocolumn,showpacs,10pt]{revtex4-1} %% Phys. Rev. A
\usepackage{amsmath,amssymb,graphicx}
\usepackage{todonotes}
\usepackage{subfigure}
\usepackage[export]{adjustbox}
\usepackage{array}
\usepackage[english]{babel}
\newcommand{\etal}{{\it et al.}}

\begin{document}
%\twocolumn[
\title{Toward the formation of crossed laser-induced periodic surface structures}
\author{Jean-Luc D\'eziel}
\author{Joey Dumont}
\author{Denis Gagnon}
\author{Louis J. Dub\'e}
\email{Louis.Dube@phy.ulaval.ca}
\affiliation{D\'epartement de physique, de g\'enie physique et d'optique \\Facult\'e des Sciences et de G\'enie, Universit\'e Laval, Qu\'ebec G1V 0A6, Canada}
\author{Sandra Messaddeq}
\author{Youn\`es Messaddeq}
\affiliation{Centre d'Optique, Photonique et Laser, Universit\'e Laval, Qu\'ebec, Qu\'ebec G1V 0A6, Canada}

\begin{abstract}
The formation of a new type of laser-induced periodic surface structures using a femtosecond pulsed laser is studied on the basis of the Sipe-Drude theory solved with a FDTD scheme. Our numerical results indicate the possibility of coexisting structures parallel \textit{and} perpendicular to the polarization of the incident light for low reduced collision frequency ($\gamma/\omega \lesssim 1/4$, where $\omega$ is the laser frequency). Moreover, these structures have a periodicity of $\Lambda \sim \lambda$ in both orientations. To explain this behavior, light-matter interaction around a single surface inhomogeneity is also studied and confirms the simultaneous presence of surface plasmon polaritons and radiation remnants in orthogonal orientations at low $\gamma/\omega$ values.
\end{abstract}

%\pacs{42.25.Bs 52.38.-r} % REPLACE WITH CORRECT PACS CODES FOR YOUR ARTICLE
%]

\maketitle

%Structure
%Introduction
%\section{The theoretical model}\label{sec:Theory}
%\section{The numerical implementation}\label{sec:FDTD}
%\section{Results of the simulations}\label{Results}
%\subsection{Formation of cross-superposed LIPSSs} \label{subsec:OldNew}
%\subsection{Collective effects of single scatterers} \label{subsec:singleScat}
%\section{Conclusions}\label{sec:Conclusions}

\section{Introduction}
%\paragraph{Introduction.}
Laser-induced periodic surface structures (LIPSSs) are the result of the strong interaction between light and matter near the ablation threshold of metals, semi-conductors or dielectrics.\cite{Birnbaum1965,Sipe1983,Young1983} The inhomogeneous absorption of energy under the target surface leads to a deeper ablation at the maxima of the electromagnetic field and the formation of wavy nanometric structures. With an increasing number of different kinds of observed structures, it has recently become one of the most direct way to study a great variety of complex light-matter interactions on various materials.\cite{Driel1982,Young1983,Bolle1992,Wu2003,Bonse2005,Couillard2007,Bonse2009,Dufft2009,Bonse2012,Hohm2012}

LIPSSs can be categorized by their orientation with respect to the polarization of light and their periodicity $\Lambda$ compared to the wavelength $\lambda$ of the incident beam. The most common structures, which have a spatial periodicity of $\Lambda \sim \lambda$, are oriented perpendicular to the polarization direction and are called low spatial frequency LIPSSs (LSFLs) or ripples. They can be observed, for instance, on strongly absorbing materials (metals or highly ionized semi-conductors/dielectrics) when $\mathrm{Re}(\tilde{n})<\mathrm{Im}(\tilde{n})$, where $\tilde{n}$ is the complex refractive index. This type of LIPSSs is referred as type-s because of the sinusoidal dependency between their periodicity and the angle of incidence of the laser beam.\cite{Young1983} On dielectrics or semi-conductors, the dominant observed structures are called high spatial frequency LIPSSs (HSFLs) when $\mathrm{Re}(\tilde{n})>\mathrm{Im}(\tilde{n})$. Their periodicity is closer to $\Lambda \sim \lambda/$Re$(\tilde{n})$ and they are oriented parallel to the polarization of the incident light. They are referred to as type-d LIPSSs (\textit{d} stands for \textit{dissident}). HSFLs with $\Lambda \ll \lambda$ oriented perpendicular to the polarization\ can also be observed and are referred to as type-r.\cite{Skolski2014} A summary of these behaviors is presented in Table \ref{behaviors}.

Various mechanisms have been proposed to explain the formation of these structures. It has been suggested that perpendicular to polarization HSFLs are caused by second harmonic generation\cite{Borowiec2003}, self-organization \cite{Reif2002} and a number of other electromagnetic explanations.\cite{Wu2003,Dufft2009,Skolski2014}
For LSFLs (type-s), it is generally accepted that they are the result of the excitation of surface plasmon polaritons (SPPs),\cite{Zayats2005,Bonse2009} which consist of trapped light at the surface of a conducting material, including ionized dielectrics. They can be produced when the incident light interacts with the surface roughness and they have a periodicity slightly smaller than $\lambda$, as for LSFLs. The effect of a single source of SPPs on laser ablation can be observed experimentally by controlling the surface roughness; for instance, by depositing a single nanoparticle on a flattened surface before the laser processing.\cite{Hecht1996,Bonse2009,Yang2014} The importance of radiation remnants has also been pointed out by Sipe \etal \cite{2Young1983,Sipe1985}in the LIPSSs formation process as non-radiative waves that can behave like SPPs on any surface, even dielectrics.

\begin{table*}
\caption{Classification of the different types of LIPSSs with their qualitative representation in the frequency domain. Dashed circles indicate where $|\vec{k}_{x,y}|=1$ and dotted circles where $|\vec{k}_{x,y}|=$ Re$(\tilde{n})$, both normalized to the norm of the incident wave number, $|\vec{k}_i|=2\pi/\lambda$. Polarization orientation of the incident light is indicated by the white arrows. \label{behaviors}}
\begin{tabular}{m{3cm} m{7cm} c}
\hline \hline
\centering \vspace{1mm} Classification & \vspace{1mm} \centering Description & Representation \\
\hline
\centering type-d & HSFLs parallel to polarization with periodicity of $\Lambda \sim \lambda/\mathrm{Re}(\tilde{n})$. Dominant in dielectrics and semi-conductors when $\mathrm{Re}(\tilde{n})>\mathrm{Im}(\tilde{n})$ due to radiation remnants close to the rough surface. & \includegraphics[valign=m,scale=0.25]{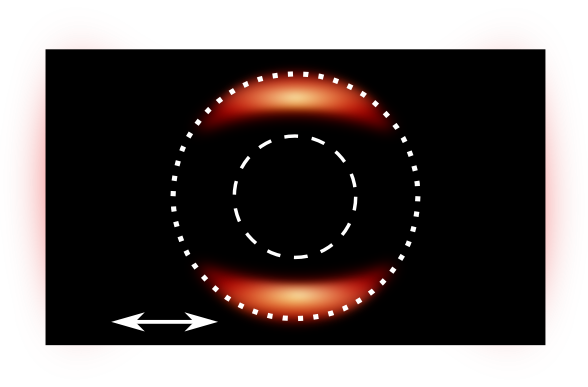} \\
\centering type-s & Structures perpendicular to polarization with periodicity of $\Lambda \sim \lambda$. Dominant in most materials when $\mathrm{Re}(\tilde{n})<\mathrm{Im}(\tilde{n})$ due to the excitation of \textit{p}-polarized SPPs around surface inhomogeneities. & \includegraphics[valign=m,scale=0.25]{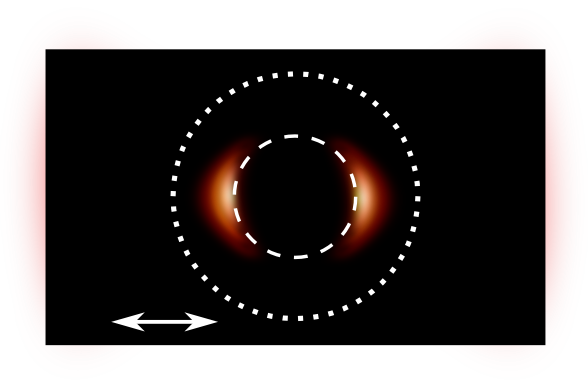} \\
\centering type-m & Novelty of the present work. Structures parallel to polarization with periodicity of $\Lambda \sim \lambda$. They occur near the crossing point of $\mathrm{Re}(\tilde{n})$ and $\mathrm{Im}(\tilde{n})$ when $\gamma/\omega$ is small. They are the results of radiation remnants and decay slowly in the material when compared with type-s behavior.  & \includegraphics[valign=m,scale=0.25]{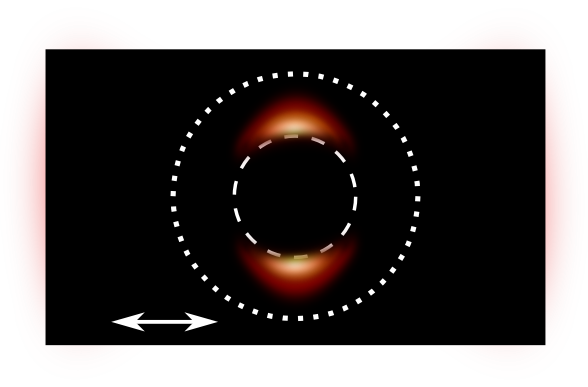} \\ \centering type-r & HSFLs perpendicular to polarization with periodicity of $\Lambda \ll \lambda$. In the FDTD approach of the Sipe-Drude theory, they are strongly roughness dependent and rapidly decay in the material.  & \includegraphics[valign=m,scale=0.25]{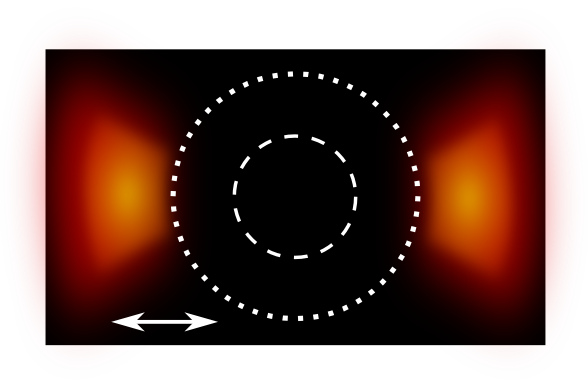} \\
\hline
\end{tabular}
\end{table*}

In this paper, we demonstrate the possibility of the formation of a new type of LIPSSs. A superposition of LSFLs oriented simultaneously in the direction perpendicular \textit{and} parallel to the polarization of incident light with $\Lambda \sim \lambda$ in both directions, resulting in structures similar to a two dimensional grid at the target's surface. We refer to these structures as crossed LIPSSs, or c-LIPSSs, which are different from other observed bidimensional structures consisting of a superposition of type-s and much smaller type-d structures.\cite{Skolski2014} c-LIPSSs result from the superposition of type-s structures and of what we will refer to as type-m structures. The \textit{m} stands for \textit{mixed} because they present characteristics of both type-s and type-d behaviors. Type-m structures are an extension of type-d features when $\mathrm{Re}(\tilde{n})$ is small, explaining their larger periodicity and parallel orientation. As for type-s structures, their periodicity remains at $\Lambda \sim \lambda$ even when $\mathrm{Re}(\tilde{n})$ is close to $0$. Also, the radiation remnants responsible for the type-m structures are similar to the type-s behavior at low $\mathrm{Im}(\tilde{n})$ values. Type-m and type-s structures coexist with similar amplitudes near the crossing point $\mathrm{Re}(\tilde{n})=\mathrm{Im}(\tilde{n})$ when the reduced collision frequency $\gamma/\omega$ taken from the Drude model\cite{Ashcroft1976} is small.

Our study is organized as follows. In Sec. \ref{sec:Theory}, we introduce the theoretical model and its constituting physical ingredients, the Sipe-Drude theory. This is followed in Sec. \ref{sec:FDTD} with the details of our numerical implementation of the finite-difference time-domain (FDTD) algorithm. The results of the simulations (over a wide range of parameters) and their interpretations are then presented in Sec. \ref{Results}. In particular, our study confirms the existence of c-LIPSSs and explain their formation process via the excitation of both SPPs and radiation remnants in orthogonal orientations. Finally, our conclusions are stated in the last section.

\section{The theoretical model}\label{sec:Theory}

%\paragraph*{The theoretical model.}
The analytical approach of choice to explain the formation of LIPSSs is the Sipe-Drude theory.\cite{Sipe1983,Bonse2009} The main idea is to solve Maxwell's equations
\begin{align}
\vec{\nabla} \times \vec{H} = \sigma \vec{E}+\epsilon_0\epsilon_r \frac{\partial \vec{E}}{\partial t}\\
\vec{\nabla} \times \vec{E} = -\mu_0 \frac{\partial \vec{H}}{\partial t}\label{maxwell2}
\end{align}
for the propagation of a plane wave through the rough surface of a material with relative complex permittivity $\tilde{\epsilon}=\epsilon'+i\epsilon''$. Conductivity is defined as $\sigma=\epsilon_0\epsilon'' \omega$ where $\omega$ is the angular frequency of the incident light and the relative (real) permittivity is expressed as $\epsilon_r=\epsilon'/\epsilon_0$ in units of the free-space permittivity $\epsilon_0$. Permeability $\mu_0$ is assumed constant. The ensuing theory predicts periodic maxima and minima of the energy profile at the surface in good agreement with experimental data\cite{Sipe1983,Young1983} under the reasonable assumption that the ablation process ejects more matter where the energy is greater. Surface roughness is confined to a thin region above the surface in which a fraction of the volume is randomly filled. The thickness of this region is realistically chosen to be significantly smaller than the incident beam wavelength.

The second ingredient consists of using the model of Drude for the complex permittivity to account for the excitation of the material,\cite{Sokolowski1999,Bonse2009}
\begin{equation}
\tilde{\epsilon}=\epsilon + \tilde{\epsilon}_{\mathrm{Drude}}= \epsilon + (1 -i \gamma/\omega) \left[ - \frac{\omega_p^2/\omega^2}{(1 + \gamma^2/\omega^2)} \right]
\label{drude}
\end{equation}
with the permittivity of the non excited material $\epsilon$, the collision frequency $\gamma = 1/\tau_D$ defined as the inverse of the Drude damping time $\tau_D$ and the plasma frequency $\omega_p$ expressed as
\begin{equation}
\omega_p = \left[ \frac{e^{2}N_e}{\epsilon_0 m^{*}_{\mathrm{opt}} m_e} \right]^{1/2} ,
\label{plasma_freq}
\end{equation}
where $e$ is the electric charge,  $N_e$ is the electronic density in the conduction band and $m^{*}_{\mathrm{opt}}m_e$ is the effective optical mass of the electron. The complex refractive index $\tilde{n}$ is then simply $\tilde{n}=\tilde{\epsilon}^{1/2}$ or, specifically,
\begin{align}
&\mathrm{Re}(\tilde{\epsilon})=\mathrm{Re}(\tilde{n}^{2})-\mathrm{Im}(\tilde{n}^{2}) \nonumber \\
&\mathrm{Im}(\tilde{\epsilon})=2\mathrm{Re}(\tilde{n})\mathrm{Im}(\tilde{n}).\label{re(epsilon)}
\end{align}
The dimensionless parameters $\omega_p/\omega$ and $\gamma/\omega$  will be used hereafter to characterize the different physical regimes. The electronic density in the conduction band can be calculated from $\omega_p/\omega$ and the laser source wavelength $\lambda$ with
\begin{equation}
N_e[10^{21}\mathrm{cm}^{-3}]=m^{*}_{\mathrm{opt}}\left( \frac{\omega_p}{\omega} \right)^{2}\frac{1.11 \cdot 10^{6}}{(\lambda [\mathrm{nm}])^{2}}.
\end{equation}

The analytical solutions of Maxwell equations under these assumptions predict,\cite{Bonse2009} for small plasma frequencies $\omega_p$ or low excitation densities $N_e$ (dielectrics of semi-conductors for which $\mathrm{Re}(\tilde{n})>\mathrm{Im}(\tilde{n})$), a dominant type-d behavior. These structures are parallel to the polarization of the incident light with periodicity $\Lambda \sim \lambda/\mathrm{Re}(\tilde{n})$. In contrast, for higher plasma frequencies, or high excitation densities $N_e$ (material with  $\mathrm{Re}(\tilde{n})<\mathrm{Im}(\tilde{n})$), this theoretical model predicts\cite{Sipe1983,Bonse2009} a dominant type-s behavior, structures oriented perpendicular the the incident light polarization and periodicity of $\Lambda \sim \lambda$. 

The analytic solutions of the Sipe-Drude theory, however, cannot properly describe the type-r behavior because they are obtained under a small frequency approximation.\cite{Skolski2012} In the frequency domain, the amplitude of the type-r behavior tends to grow indefinitely for larger frequencies. In this regime, a numerical solution easily solves this difficulty.

The effects of the parameter $\gamma/\omega$ are still largly unexplored in this theory. Most theoretical studies\cite{Bonse2009,Bonse2010,Skolski2012,Skolski2014} have restricted their efforts to a laser source of $\lambda=800$ nm ($\omega=2.35 \cdot 10^{15}$ s$^{-1}$, $T=2\pi/\omega=2.67$ fs) incident on a silicon target ($\tau_D = 1.1$ fs), corresponding to a reduced collision frequency $\gamma/\omega \sim 0.39$. Since this parameter is linked to the relaxation of the electron density and is expected to vary for different materials of interest,\cite{Johnson1972,Ashcroft1976} we will scan different values of $\gamma/\omega$ in order for our computations to access a variety of target materials.

\section{The numerical implementation}\label{sec:FDTD}
%\paragraph*{The numerical implementation.}
We follow the lead of Skolski \etal \cite{Skolski2012} who were recently successful in numerically modeling the formation of LIPSSs 
with a FDTD solver.\cite{Yee1966,Taflove2005} We thus use this method for its versatility and adapt it to explain the formation of the c-LIPSSs. The details of our geometry and its spatio-temporal discretization are presented next.

\begin{figure}[t]
\centering
\includegraphics[scale=0.82]{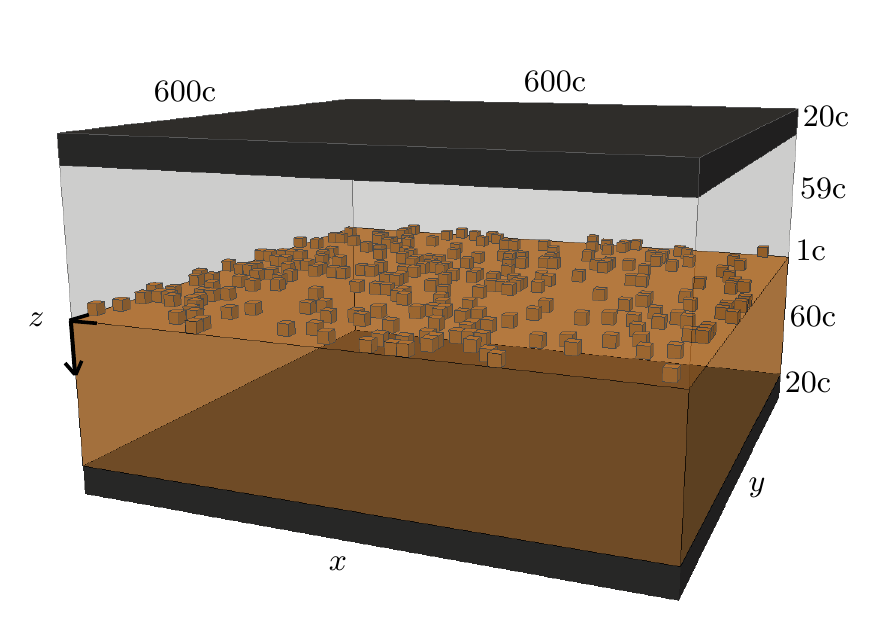}
\caption{(Color online) Schematic representation of the geometry used in the FDTD simulations. The black zones re\-present PMLs, the upper grey zone consists of vacuum and the lower orange zone is the material. The plane wave source is the bottom layer of the upper PML and propagates in the $+z$ direction towards the material. Note that vertical walls also have PMLs of 20 cells wide (not shown).} \label{geometry}
\end{figure}

\begin{figure}[h]
\centering
\includegraphics[scale=0.38]{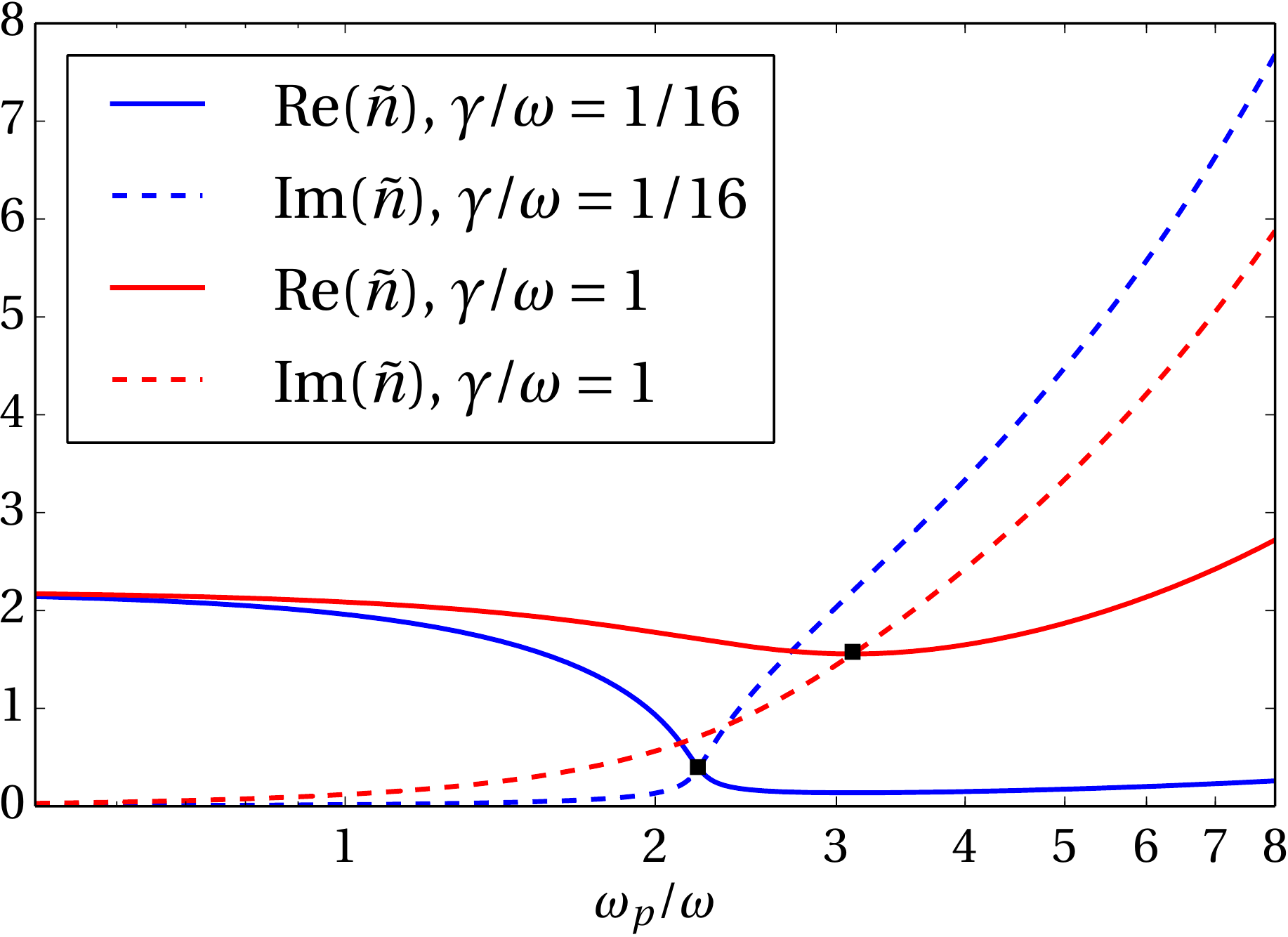}
\caption{(Color online) Real (full lines) and imaginary (dashed lines) parts of the refractive index $\tilde{n}=\tilde{\epsilon}^{1/2}$. The non-excited material has a purely real permittivity of $\epsilon=4.84$ (index $n=2.2$) and the excited permittivity $\tilde{\epsilon}$ follows Eq. \eqref{drude}. The crossing points, $(\omega_p/\omega)_c$, are indicated with black dots. Their numerical values are 2.2 and 3.11 for $\gamma/\omega=1/16$ and 1 respectively.}\label{index}
\end{figure}

\begin{figure*}
\centering
\includegraphics[scale=1.0]{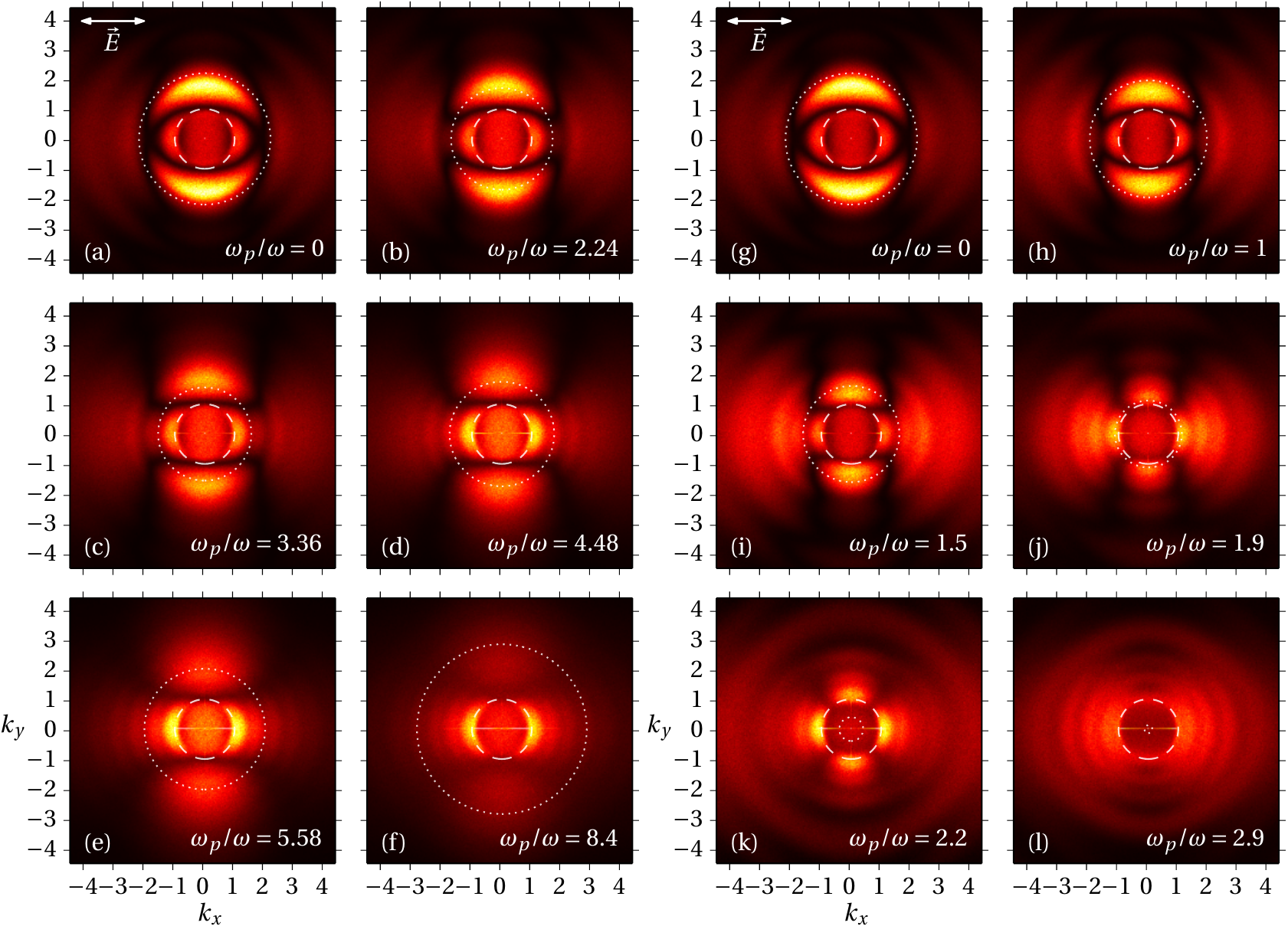}
\caption{(Color online) Fourier transforms amplitudes (linear color scale with arbitrary normalization) of $\langle|\vec{E}|^{2}\rangle_{x,y}$ for $\gamma/\omega=1$ in subfigures (a) - (f),  $\gamma/\omega=1/16$ in subfigures (g) - (l) and the indicated value of $\omega_p/\omega$. The wave numbers $\vec{k}_{x,y}$ are normalized to the norm of the incident wave number, $|\vec{k}_i|=2\pi/\lambda$. Dashed circles indicate where $|\vec{k}_{x,y}|=1$ and dotted circles where $|\vec{k}_{x,y}|=~$Re$(\tilde{n})$. Polarization of the incident light is along the $x$ axis. }\label{fourier}
\end{figure*}

Each simulation is performed on a space domain of 32 wavelengths in the $x$ and $y$ directions and 8 wavelengths in the $z$ direction with a discretization of 20 cells per wavelength: 
$\delta_{x,y,z}=\delta= \lambda/20$ for a total simulation volume of $[X_S \times Y_S \times Z_S]$ with $X_S=Y_S= 640\  \delta$ and $Z_S= 160\  \delta$.
Time increments have to be carefully chosen in order to ensure stability over the entire simulation. The stability condition for the FDTD method is 
\begin{equation}
\delta_{t,\mathrm{max}}=\frac{1}{c}\left[ \frac{1}{\delta x^{2}}+ \frac{1}{\delta y^{2}}+ \frac{1}{\delta z^{2}} \right]^{-1/2},
\end{equation}
or $\delta_{t,\mathrm{max}}=T/(20\sqrt{3})$ with our spatial discretization. We therefore use a time discretization $\delta_t<\delta_{t,\mathrm{max}}$ of 40 time steps per optical cycle, $T=2 \pi/\omega$, and the simulations last for 10 cycles: $\delta_t= T/40$ for a total time interval of $T_S= 10\ T = 400\ \delta_t$. Furthermore, absorbing material, represented by perfectly matched layers (PMLs) \cite{Johnson2010} of 20 cells wide are positioned at each boundary of the domain. Also, to reduce unwanted boundary effects, only a central 28x28 wavelengths sub-domain is used in the subsequent analysis. The surface rugosity is modeled by a small region $-\delta < z \leqslant 0$ with a surface content following a random binary function. Specifically, on the surface, one random computational cell out of ten is occupied with material, ensuring an approximate filling factor of 10\%. The geometry and the spatial discretization are displayed in Figure \ref{geometry}.

A plane wave is propagated on the spatio-temporal grid in the $+z$ direction with a linear polarization along the $x$ axis. The region $z\geqslant 0$ consists of a material with permittivity $\tilde{\epsilon}=\epsilon + \tilde{\epsilon}_{\mathrm{Drude}}$ where the non-excited material contributes the constant $\epsilon$. The same $\tilde{\epsilon}$ accounts for the surface material. The values of the refractive index, $\tilde{n}=\tilde{\epsilon}^{1/2}$, used in this work are shown in Figure \ref{index}. 

Some features in the figure are worth noting. The region where $\mathrm{Re}(\tilde{n})=\mathrm{Im}(\tilde{n})$ will be of special importance, since it defines the boundary between different physical regimes, as we will see later. Clearly this crossing point is the position where $\epsilon'$ changes sign, precisely when $(\omega_p/\omega)_c=\epsilon^{1/2}(1+\gamma^{2}/\omega^{2})^{1/2}$. Furthermore, as seen from equation \eqref{re(epsilon)}, at the point of intersection of the curves, $\epsilon''=(\gamma/\omega)\epsilon^{1/2}$ and the value at $(\omega_p/\omega)_c$ varies as $\mathrm{Re}(\tilde{n})_c=\mathrm{Im}(\tilde{n})_c=(\gamma/\omega)^{1/2}\epsilon^{1/4}/\sqrt{2}$.

\begin{figure*}
\centering
\includegraphics[scale=1.1]{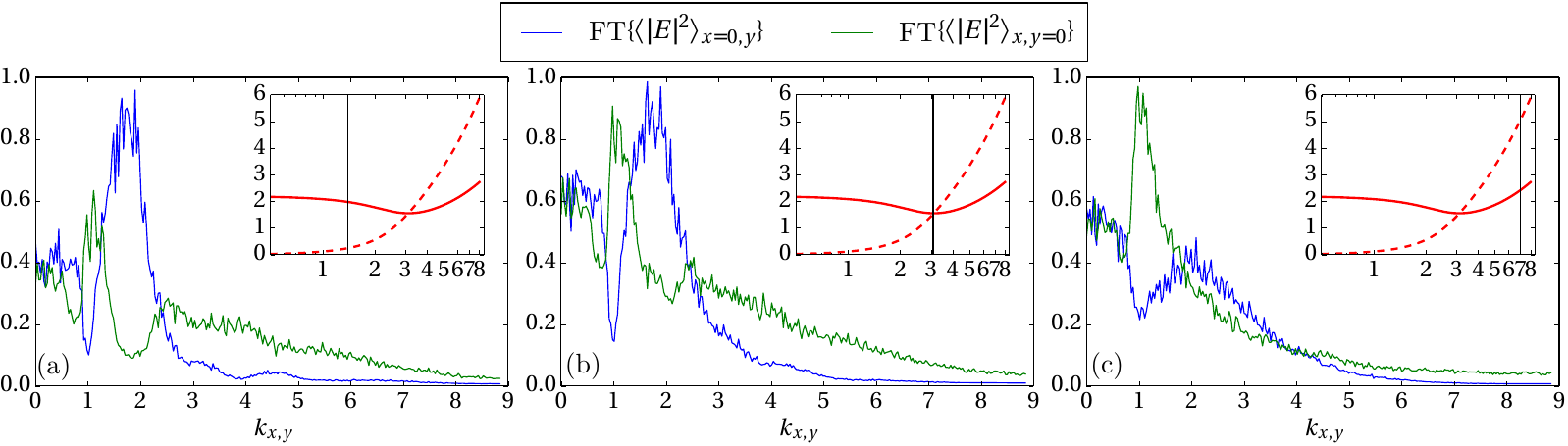}
\caption{(Color online) Evolution of the Fourier transforms along both $k_x=0$ and $k_y=0$ axes for $\gamma/\omega=1$ as $\omega_p/\omega$ varies from below (a) to above (c) $(\omega_p/\omega)_c=3.11$ (b). Amplitudes of the Fourier transforms are normalized to maintain the maximum value at $1$, but their relative amplitude is preserved. The insets display $\mathrm{Re}(\tilde{n})$ with a full line and $\mathrm{Im}(\tilde{n})$ with a dashed line together with a vertical line indicating the current $\omega_p/\omega$ value.}\label{gamma1frames}
\end{figure*}

To obtain Figure \ref{fourier}, we perform $50$ simulations for each pair ($\omega_p/\omega$, $\gamma/\omega$) with a reshuffled rugosity and apply a bidimensional Fourier transform on
the function $\langle|\vec{E}|^{2}\rangle_{x,y}$ which is the squared electric field evaluated at $z= +\delta$ averaged over the second half of the time domain, i.e.
\begin{equation}
\langle |\vec{E}|^2 \rangle_{x,y} = \frac{2}{T_S} \int_{T_S/2}^{T_S} |\vec{E}(x,y,z=+\delta;t)|^2 dt.
\label{moyenne}
\end{equation}
Finally, we average over the $50$ Fourier transforms for each pair of parameters to obtain smoother results. 

Three points should be mentioned. First, the time average performed in equation \eqref{moyenne} extends over an integer number of complete optical cycles, $5$ with our temporal discretization. Second, we evaluate the solutions at $z=+\delta$ (corresponding to $40$ nm for a $800$ nm laser source) in order to minimize any effects related to the specific representation of the surface roughness. Something that the Sipe equations cannot achieve, as they are restricted to the plane $z=0$. Third, the possibility to investigate below the surface is then a notable advantage of the FDTD approach, providing solutions over the entire simulation volume.

\section{Results of the simulations}\label{Results}

We have carried out an exhaustive scan of parameter space $(\omega_p/\omega,\gamma/\omega)$ for $\omega_p/\omega \in [0,10]$ and $\gamma/\omega \in [0,4]$ with the model described in Sec. \ref{sec:FDTD}. For convenience, the non-excited permittivity $\epsilon$ has been kept fixed to the numerical value 4.84 ($n=2.2$) for the entire sets of calculations. Our exploration has allowed us to isolate two separate regimes of different qualitative behaviors. The \emph{approximate} boundary between them is found at $\gamma/\omega \sim 1/4$. This value will vary somewhat with $\epsilon$ since the two regimes are affected by the position of $(\omega_p/\omega)_c \propto \epsilon^{1/2}$ delimiting itself dynamical changes within each regime.

To appreciate the qualitative features of the two regimes, we have chosen to display two representative series of calculations with the choices $\gamma/\omega=1$ and $\gamma/\omega=1/16$.

\subsection{Formation of c-LIPSSs} \label{subsec:OldNew}

The case $\gamma/\omega=1$ (see Figures \ref{fourier}(a) - (f)) shows the more common transition from the type-d behavior to the type-s behavior as $\omega_p/\omega$ increases. The real and imaginary parts of the refractive index cross at $(\omega_p/\omega)_c = 3.11$ roughly corresponding to Figure \ref{fourier}(c). At higher plasma frequencies, Figures \ref{fourier}(d) - (f), perpendicular structures (type-s behavior) become gradually dominant, since it is the region where $\mathrm{Re}(\tilde{n})<\mathrm{Im}(\tilde{n})$ and SPPs can be excited. This transition has been extensively studied before with the parameters $\lambda = 800$ nm and $\tau_D = 1.1$ fs, corresponding to $\gamma/\omega = 0.39$.\cite{Bonse2009,Bonse2010,Skolski2012,Skolski2014,Yang2014} The evolution of the Fourier transforms along both $k_x=0$ and $k_y=0$ axes as $\omega_p/\omega$ grows for $\gamma/\omega=1$ is shown in Figure \ref{gamma1frames}. These are two orthogonal slices of the correspong two-dimensional calculations of Figures \ref{fourier}(a) - (f). The amplitudes of the Fourier transforms are normalized to maintain the maximum value at $1$, but their relative amplitude is preserved. The transition from type-d (Figure \ref{gamma1frames}(a)) to type-s (Figure \ref{gamma1frames}(c)) behavior arises very close to $(\omega_p/\omega)_c$ (Figure \ref{gamma1frames}(b)).

\begin{figure*}
\centering
\includegraphics[scale=1.1]{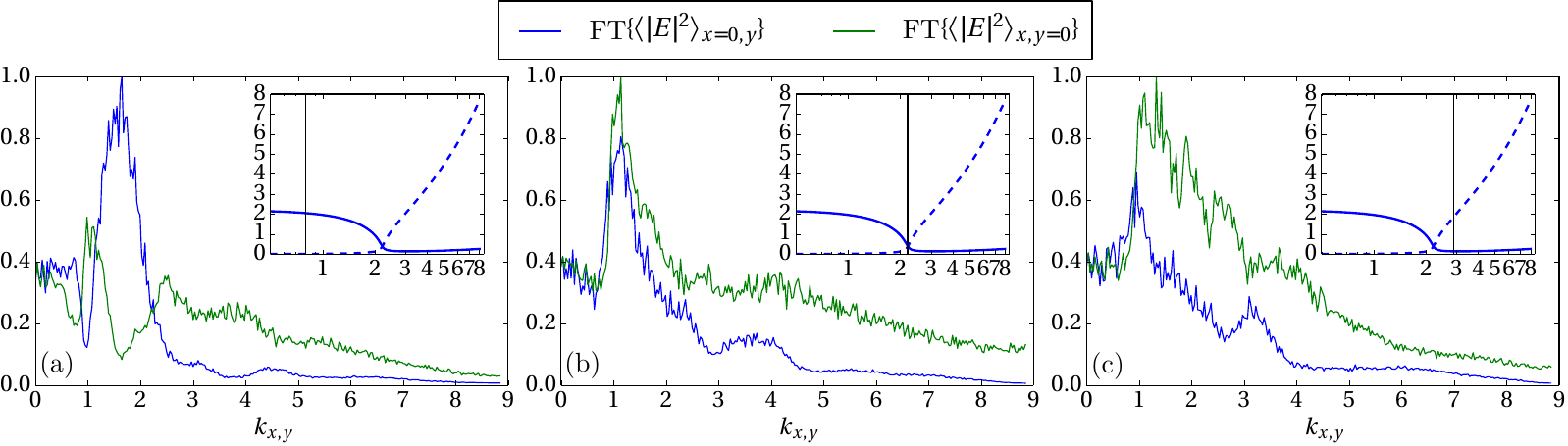}
\caption{(Color online) Evolution of the Fourier transforms along both $k_x=0$ and $k_y=0$ axes for $\gamma/\omega=1/16$ as $\omega_p/\omega$ varies from below (a) to above (c) $(\omega_p/\omega)_c=2.2$ (b). Amplitudes of the Fourier transforms are normalized to maintain the maximum value at $1$, but their relative amplitude is preserved. The insets display $\mathrm{Re}(\tilde{n})$ with a full line and $\mathrm{Im}(\tilde{n})$ with a dashed line together with a vertical line indicating the current $\omega_p/\omega$ value.}\label{gamma2frames}
\end{figure*}

\begin{figure}
\centering
\hspace*{-0.5cm}
\includegraphics[scale=0.95]{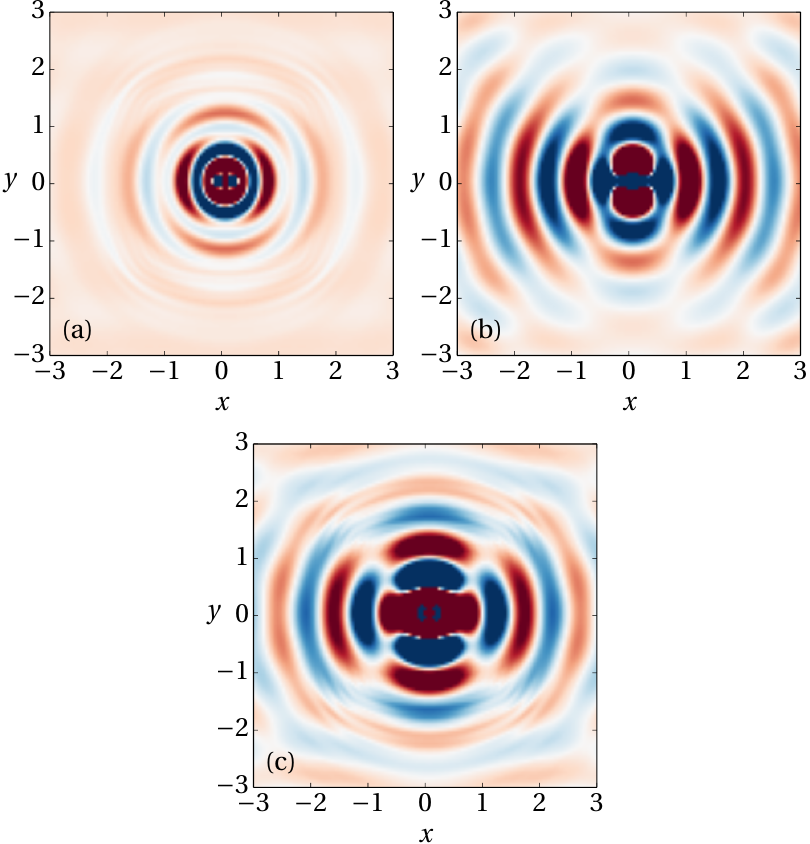}
\caption{(Color online) Spatial domain solutions of $\langle|\vec{E}|^{2}\rangle_{x,y}$ for (a) $(\gamma/\omega,\omega_p/\omega)=(1,0)$, (b) $(\gamma/\omega,\omega_p/\omega)=(1,8.4)$ and (c) - (d) $(\gamma/\omega,\omega_p/\omega)=(1/16,1.9)$. Only one inhomogeneity is positioned at the center of the surface. Polarization of the incident light is along the $x$ axis. Color scaling is arbitrary with blue for minimal values and red for maximal values.}\label{singlePara}
\end{figure}

For smaller values of $\gamma/\omega$, below $\sim 1/4$, we find the same type-d behavior for low values of $\omega_p/\omega$ and as we get closer to $(\omega_p/\omega)_c=2.2$, the type-s behavior gradually becomes dominant, in addition to other structures parallel to the incident light polarization with periodicity of $\Lambda \sim \lambda$, the type-m behavior. The simultaneous presence of type-s and type-m behaviors with a similar amplitude forms the c-LIPSSs. Figures \ref{fourier}(g) - (l) illustrates this behavior for $\gamma/\omega=1/16$. The transition occurs in the region $\omega_p/\omega \sim 1.9-2.2$ (see Figure \ref{fourier}(k)), where the real and imaginary parts of the refractive index are nearly equal and small (see Figure \ref{index}). The evolution of the Fourier transforms along axes $k_x=0$ and $k_y=0$ as $\omega_p/\omega$ varies, with $\gamma/\omega=1/16$ is shown in Figure \ref{gamma2frames}. Again, the relative amplitude of the two Fourier transforms is preserved. The initial maxima (Figure \ref{gamma2frames}(a)) correspond to type-d behavior at $\Lambda \sim \lambda/\mathrm{Re}(\tilde{n})$ which gradually shifts to $\Lambda \sim \lambda$ (type-m) as $\omega_p/\omega$ grows. Near $(\omega_p/\omega)_c$, type-m and type-s coexist with similar amplitudes (Figure \ref{gamma2frames}(b)). It is interesting to see that the type-m maxima is present and remains at $\Lambda \sim \lambda$ even when $\mathrm{Re}(\tilde{n})<\mathrm{Im}(\tilde{n})$ and $\mathrm{Re}(\tilde{n})$ is close to $0$. At higher plama frequencies, type-m behavior is still present, but type-s becomes dominant (Figure \ref{gamma2frames}(c)).

In Sec. \ref{subsec:singleScat}, we investigate the effects of a single source of radiation remnants or SPPs by replacing the surface rugosity with one single inhomogeneity.

\subsection{Collective effects of single scatterers} \label{subsec:singleScat}

Figure \ref{fourier} shows results of incident light interacting with a large number of surface inhomogeneities or scatterers. We can reduce the problem to a single scatterer to isolate the effects on the field caused by one inhomogeneity for different values of ($\gamma/\omega$,$\omega_p/\omega$). We have therefore performed a number of simulations using the same method as described in Sec. \ref{sec:FDTD}, but with the scatterer positioned at the center of the surface and a reduced domain of $X_S=Y_S= 320\  \delta $ and $Z_S= 160\  \delta$. This inhomogeneity has the same size (one cell) and properties as before. These simulations serve two purposes: first, to confirm that we obtain the same pattern as the experimental findings\cite{Bonse2009,Yang2014} for parameters leading to the type-s behavior and second, to compare with the effects of the radiation remnants around one inhomogeneity with the parameters that should lead to c-LIPSSs.

The resulting fields $\langle |\vec{E}|^2 \rangle_{x,y}$ in the space domain are shown in Figure \ref{singlePara}. For the pair of parameters $(\gamma/\omega,\omega_p/\omega)=(1,0)$, we obtain Figure \ref{singlePara}(a) where no SPP is expected. For the pair of parameters $(\gamma/\omega,\omega_p/\omega)=(1,8.4)$, on Figure \ref{singlePara}(b), we see the result of interaction between SPPs propagating along the $x$ axis (\textit{p}-polarized SPPs) and incident light, a pattern similar to the experimental results of laser processed surfaces around single nanoparticles.\cite{Bonse2009,Yang2014} With the pair of parameters $(\gamma/\omega,\omega_p/\omega)=(1/16,1.9)$ that should lead to the formation of c-LIPSSs, we still obtain, as shown in Figure \ref{singlePara}(c), oscillations along the $x$ axis responsible for type-s behavior in addition to \textit{s}-polarized radiation remnants resulting in oscillations along the $y$ axis, themselves responsible for type-m behavior. They are \textit{s}-polarized because they propagate in the $y$ direction with nearly all of their energy is in the $x$ component of the electric field. These latter excitations cannot be SPPs since \textit{s}-polarized SPPs can only propagate on metamaterials which exhibit a negative permeability.\cite{BaoRong2010} This can be shown by solving Maxwell's equations in a 2D domain, or the equivalent Helmholtz equation $(\vec{\nabla}^{2}+\vec{k}^{2})\vec{E}=0$, near the interface. In the $(y,z)$ plane, a \textit{s}-polarized interface mode between two media is described by
\[
 \vec{E}(y,z,t) =
  \begin{cases}
   \hat{x}E_1\exp(ik_{y1}y+\alpha_{1}z-i\omega t), & z < 0, \\
   \hat{x}E_2\exp(ik_{y2}y-\alpha_{2}z-i\omega t),  & z > 0,
  \end{cases}
\]
where the indices $1$ and $2$ denote the two different media above and below the surface respectively. The wavenumbers $(k_{y1},k_{y2})\geq 0$ represent oscillations in the $y$ direction and the purely real wavenumbers $(\alpha_{1},\alpha_{2})\geq 0$  describe the exponential field decay away from the surface. Using equation \eqref{maxwell2}, we find the corresponding magnetic field
\begin{flalign*}
&\quad\vec{H}(y,z,t) =& \\
  & \begin{cases}
   -\left[ \hat{y}\frac{E_1\alpha_{1}}{i\omega \mu_1}+\hat{z}\frac{E_1k_{y1}}{\omega \mu_1}\right]\exp(ik_{y1}y+\alpha_{1}z-i\omega t), & z < 0, \\ 
    \left[ \hat{y}\frac{E_2\alpha_{2}}{i\omega \mu_2}-\hat{z}\frac{E_2k_{y2}}{\omega \mu_2}\right]\exp(ik_{y2}y-\alpha_{2}z-i\omega t), & z > 0,
  \end{cases}&
\end{flalign*}
with the permeabilities $\mu_1$ and $\mu_2$. The phase matching condition is obtained by equating the tangential fields at the interface $z=0$,
\begin{equation}
\frac{\alpha_{1}}{\mu_1}+\frac{\alpha_{2}}{\mu_2}=0.
\end{equation}
Phase matching is then possible when $\mu_1$ and $\mu_2$ have opposite signs. But since our simulations use constant permeabilities $\mu_1=\mu_2=\mu_0$, the phase matching condition can only be satisfied for $\alpha_{1}=\alpha_{2}=0$ resulting in \textit{s}-polarized excitations along the $y$ axis not confined near the surface and therefore, not SPPs. This result is consistent with the fact that these excitations occur where the imaginary part of the refractive index is still small implying a nearly lossless material. This can be further verified by looking at depth $z$ below the surface. 

\begin{figure}
\centering
\hspace*{-0.5cm}
\includegraphics[scale=0.9]{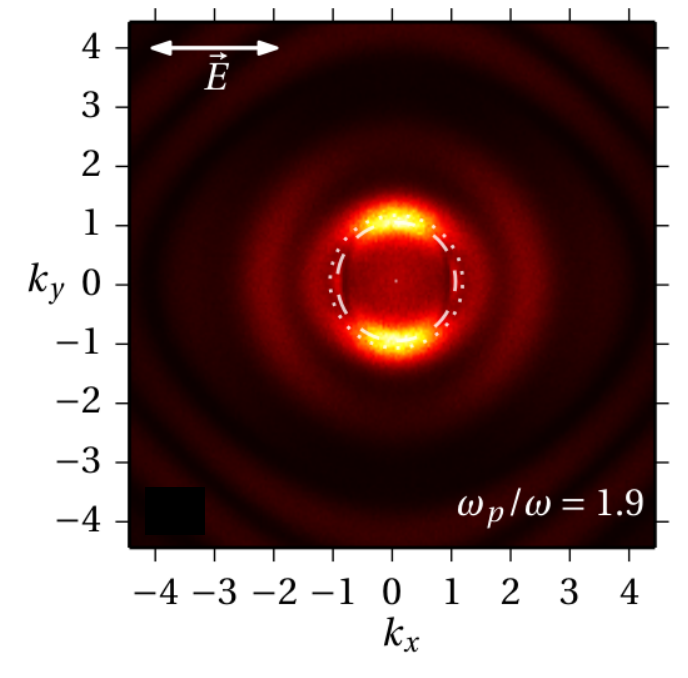}
\caption{(Color online) Fourier transforms amplitude (linear color scale with arbitrary normalization) of $\langle|\vec{E}|^{2}\rangle_{x,y}$ at $z=5\delta$ for $(\gamma/\omega,\omega_p/\omega)=(1/16,1.9)$. The wave numbers $\vec{k}_{x,y}$ are normalized to the norm of the incident wave number, $|\vec{k}_i|=2\pi/\lambda$. Dashed circle indicates where $|\vec{k}_{x,y}|=1$ and dotted circle where $|\vec{k}_{x,y}|=~$Re$(\tilde{n})$. Polarization of the incident light is along the $x$ axis. }\label{typem}
\end{figure}

With the complete rugosity simulations, as in Sec. \ref{subsec:OldNew}, and the same parameters as Figure \ref{fourier}(j), $(\gamma/\omega,\omega_p/\omega)=(1/16,1.9)$, we observe a strong attenuation of type-r and type-s structures beyond $z>3\delta$ together with an increasingly dominant type-m structures as we look deeper into the bulk. Figure \ref{typem} shows the remaining field amplitude at $z=5\delta$.

\section{Conclusions}\label{sec:Conclusions}

The formation of c-LIPSSs has been demonstrated using the FDTD version of the Sipe-Drude theory. We have found that structures parallel to the polarization direction with periodicity $\Lambda \sim \lambda$, or type-m, appear together with type-s structures for $\gamma/\omega \lesssim 1/4$ near $(\omega_p/\omega)_c$, where the real part of the permittivity vanishes. Type-m structures are caused by the presence of radiation remnants produced by the interaction between the incident light and the surface rugosity. We also find that these structures decay slowly in the bulk compared to type-s, meaning that it could be possible to grow exclusively type-m structures with deeper ablation, in less resistant materials for instance. Perhaps the most interesting feature however is that type-m and type-s structures have similar amplitudes closer to the surface. This may open the possibility to grow c-LIPSSs. The next step of this study will be the implementation of an inter-pulse feedback mechanism\cite{Skolski2014} where we could see c-LIPSSs grow and acquire better definition and stability from one pulse to the next. We also have initiated an experimental study on glassy materials and we hope to report our findings shortly in a separate contribution.

The authors acknowledge the financial support form the Canada Excellence Research Chair in Photonics Innovations of Y. Messaddeq and the Natural Sciences and Engineering Research Council of Canada (NSERC). We also acknowledge computational resources from Calcul Qu\'ebec and the free software project Meep.\cite{Meep2010}

\bibliography{lipss2014}

\end{document}